\UseRawInputEncoding
\documentclass[twocolumn]{aastex61}

\newcommand\aastex{AAS\TeX}

\newcommand{\aips}{$\mathcal{AIPS}$\,}

\newcommand{\Fig}[1]{\hyperref[#1]{Figure~\ref*{#1}}}
\newcommand{\Tab}[1]{\hyperref[#1]{Table~\ref*{#1}}}
\newcommand{\Sec}[1]{\hyperref[#1]{Section~\ref*{#1}}}
\newcommand{\Eq}[1]{\hyperref[#1]{Equation~\ref*{#1}}}
\newcommand{\github}{https://github.com/lucasjord}
\usepackage{savesym}

\usepackage{amsmath}
\restoresymbol{SIX}{tablenum}
\usepackage{hyperref}
\usepackage{booktabs}
\usepackage{comment}
\usepackage{ulem}
\usepackage{tikz}

\shorttitle{\aastex\ Inverse Multiview Calibration I}
\shortauthors{Hyland et al.}

\newcommand{\degrees}{$^\circ$}
\newcommand{\utas}{School of Natural Sciences, University of Tasmania, Private Bag 37, Hobart, Tasmania 7001, Australia}
\newcommand{\jive}{Joint Institute for VLBI ERIC, Oude Hoogeveensedijk 4, 7991PD Dwingeloo, Netherlands}
\newcommand{\ljhemail}{\url{Lucas.Hyland@utas.edu.au}}

\begin{document}
	
	\title{Inverse Multiview I: Multi--Calibrator inverse phase referencing for Microarcsecond VLBI Astrometry} 
	
\author[0000-0002-4783-6679]{L. J. Hyland}\affil{\utas}\ljhemail
\author[0000-0001-7223-754X]{M. J. Reid}\affil{Center for Astrophysics ∣ Harvard \& Smithsonian, Cambridge, MA 02138, USA}
\author[0000-0002-1363-5457]{S. P. Ellingsen}\affil{\utas}
\author[0000-0003-4871-9535]{M. J. Rioja}\affil{CSIRO Astronomy and Space Science, PO Box 1130, Bentley WA 6102, Australia}
\affil{ICRAR, M468, The University of Western Australia, 35 Stirling Hwy, Crawley, Western Australia, 6009}
\affil{Observatorio Astron\'omico Nacional (IGN), Alfonso XII, 3 y 5, 28014 Madrid, Spain}

\author[0000-0003-0392-3604]{R.  Dodson}\affil{ICRAR, M468, The University of Western Australia, 35 Stirling Hwy, Crawley, Western Australia, 6009}
\author[0000-0002-5526-990X]{G. Orosz}\affil{\utas}\affil{\jive}

\author[0000-0001-5862-4834]{C. R. Masson}\affil{Center for Astrophysics ∣ Harvard \& Smithsonian, Cambridge, MA 02138, USA}
\author[0000-0002-0233-6937]{J. M. McCallum}\affil{\utas}	

\newcommand{\LJH}[1]{{\color{orange} \textbf{$^\mathrm{LJH}$} #1}}
\newcommand{\SPE}[1]{{\color{blue} \textbf{$^\mathrm{SPE}$} #1}}
\newcommand{\MJR}[1]{{\color{red} \textbf{$^\mathrm{Mark}$} #1}}
\newcommand{\RD}[1]{{\color{green} \textbf{$^\mathrm{RD}$} #1}}
\newcommand{\MR}[1]{{\color{yellow} \textbf{$^\mathrm{Maria}$} #1}}
\newcommand{\JMC}[1]{{\color{magenta} \textbf{$^\mathrm{JMc}$} #1}}
\newcommand{\GO}[1]{{\color{cyan} \textbf{$^\mathrm{GO}$} #1}}
\newcommand{\red}[1]{#1}
	
\begin{abstract}
	Very Long Baseline Interferometry (VLBI) astrometry is a well established technique for achieving $\pm10~\mu$as parallax accuracies at frequencies well above 10~GHz.  At lower frequencies, uncompensated interferometer delays associated with the ionosphere play the dominant role in limiting the astrometric accuracy.  Multiview is a novel VLBI calibration method, which uses observations of multiple quasars to accurately model and remove time-variable, directional-dependent changes to the interferometer delay.  Here we extend the Multiview technique by phase referencing data to the target source (``inverse Multiview") and test its performance. Multiple observations with a four-antenna VLBI array operating at 8.3~GHz show single-epoch astrometric accuracies near $20~\mu$as for target-reference quasar separations up to about 7 degrees.  This represents an improvement in astrometric accuracy by up to an order of magnitude compared to standard phase referencing.
\end{abstract}
	

\section{Introduction} \label{sec:intro}
	High accuracy astrometry at radio frequencies provides fundamental information for many fields of astronomy and astrophysics \citep{ReidHonma2014}.  Very Long Baseline Interferometry has provided trigonometric parallaxes with accuracies of $10~\mu$as or better for masers associated with massive young stars throughout the Milky Way and approaching this accuracy for evolved stars \citep[e.g.][]{Reid2019,vera2020}.  In addition, parallaxes for X-ray binaries \citep[e.g.][]{MillerJones2021} and pulsars \citep[e.g.][]{Deller2019} have been critical to characterizing these sources and, in some cases, using them to test General Relativity.  Finally, measurements of extra-galactic proper motions approaching $\sim1~\mu$as~y$^{-1}$ accuracy have been accomplished \citep[e.g][]{Brunthaler2005}.  The most accurate radio frequency astrometry has been achieved at observing frequencies above $\sim10$~GHz where ionospheric delay errors, which scale with observing frequency as $\nu^{-2}$, are typically small.  However, below this frequency, these delay errors become the dominant source of astrometric error, and improving astrometric accuracy at frequencies below $\sim10$~GHz requires new approaches to measure and remove ionospheric delays \citep{RiojaDodson2020}

	The electron density in the ionosphere varies with a strong diurnal signature above a location on Earth.  Since the dominant source of ionization is solar ultra-violet radiation, the highest electron densities track the sub-solar point and this can produce strong gradients in the electron distribution both in Geographic longitude and latitude.   Models of the total electron content (TEC) as a function of time and location on the Earth can be used to partially correct for propagation delays through the ionosphere, but these models can have uncertainties of 20\% or more \citep{WalkerChatterjee1999}.	The result is that residual phase-delays of $\sim0.1$~nsec can be present in radio interferometer data at 8.3~GHz.  Importantly, these phase-delays can have significant and long-lived gradients (``ionospheric wedges") across $\sim10^\circ$ of sky, which can degrade the accuracy of relative position measurements between nearby sources \citep[e.g.][]{Reid2017}.

	Multiview is a novel approach for calibration of Very Long Baseline Interferometric (VLBI) observations designed to achieve the highest possible astrometric accuracy, particularly at frequencies below about 10~GHz where ionospheric effects are the dominant source of position errors.  The core idea of Multiview is that observations of multiple calibrators, which surround the target on the sky, allow for removal of directional, time-variable residual phase-delays from the calibrators to the target, using 2D spatial and temporal interpolation in the visibility domain.
	Initial trials of Multiview calibration, called ‘cluster-cluster’ phase referencing \citep{Rioja1997,Rioja2002}, involved simultaneous observations of a target and multiple calibrators by utilising multiple telescopes at a single site.  This method showed promise in removing residual ionospheric delays which plague low-frequency astrometry.  However, the availability of multiple telescopes at many sites is extremely limited.

	The next iteration of Multiview used only a single telescope at each site and source switching \citep{Rioja2017}.  Observations at 1.6~GHz were conducted using the Very Long Baseline Array (VLBA) and structured with three calibrators ($C_1,C_2,C_3$) surrounding a target OH maser ($T$) with the observing sequence:
	\begin{equation}
		C_1,C_2,C_3,T,C_1\dots
	\end{equation}
	We refer to this approach as \red{direct} Multiview.  A critical requirement is that all sources have to be observed within the atmospheric coherence time.  Tests showed that excellent results were achieved when one entire sequence was completed in 5 minutes at 1.6~GHz \citep{Rioja2017}.

    \citet{Reid2017} investigate fitting \red{and removing a single `positional' gradient to the measured positions of} multiple calibrators after these had been phase-referenced to the target and imaged \red{at each epoch}. \red{As this fitting was done in the image domain,} we refer to this approach as `image-based' Multiview (imMV), and it has been shown to improve astrometric accuracy at 6.7~GHz \citep{Sakai2019,Zhang2019}.

	Standard phase-referencing \red{(PR)} for VLBI observations involves `nodding' all telescopes between a calibrator and a target, measuring the phase on the calibrator and transferring it to the target \citep{Alef1986,BeasleyConway1995}. Inverse phase referencing (iPR) is commonly used for astrometry of astrophysical maser sources when the target is strong and the calibrator may be weak. Here the phase of the target is transferred to the calibrator, and ultimately the measured offset position of the calibrator is used to infer that of the target. \red{In PR/iPR (or imMV)} one \red{will often} observe a sequence of $N$ calibrators as
	\begin{equation}
		T,C_1,T,C_1,T,\dots T,C_N,T,C_N,T\dots
		\label{eq:iprnodding}
	\end{equation} 
	\red{thereby allowing either iPR or PR to be used (e.g. if the target is weaker than expected).} \red{The main benefit of the iPR technique for Multiview applications is that it} only \red{requires} that adjacent observations of the target (e.g. $T,C1,T$) are spaced by less than the coherence time of the atmosphere.  This is especially valuable at \red{frequencies around 7~GHz}, where coherence times are typically a factor of \red{5} shorter than at \red{1.4~GHz}.

	In this paper, we introduce and test ``inverse Multiview" (iMV), a combination of iPR and Multiview.  In order to test iMV we selected strong `target' quasars surrounded by compact extragalactic radio sources acting as calibrators, and observed them in multiple sessions over 4 months. By using the target as the phase-reference, we tested the positional repeatability after application of normal iPR, iMV and imMV, and compared the results.  Also, by using calibrators that surrounded the target at increasingly larger separations, we could evaluate how far separated from the target one can use calibrators without violating the assumption of planar ionospheric wedges.

\section{Observations \& Data Reduction} \label{sec:obsnreduction}
	\begin{table*}[ht]
		\centering
		\caption{Target and calibrator positions, separations and flux densities. \textbf{Columns:} Target (1) and calibrator (2) names, correlated positions in right ascension (3) and declination (4), calibrator-target offset in right ascension/East-West (5) and declination/North-South (6), total separation (7), catalogue 8.3~GHz flux density (8), average synthesised image integrated intensity (9).}
	    \label{tab:sources}	
		\begin{tabular}{llccrrlcc} \hline
			\multicolumn{2}{c}{\bf Source} & \multicolumn{1}{c}{\bf R.A.} & \multicolumn{1}{c}{\bf Dec.} & \multicolumn{3}{c}{\bf Separation} & \multicolumn{2}{c}{\bf Flux} \\
			\multicolumn{1}{c}{\bf Target} & \multicolumn{1}{c}{\bf Calibrators} & \multicolumn{1}{c}{(J2000)} & \multicolumn{1}{c}{(J2000)} & \multicolumn{1}{c}{$\Delta\alpha\times$} & \multicolumn{1}{c}{$\Delta\delta$} & \multicolumn{1}{c}{$\theta_\mathrm{sep}$} & \multicolumn{2}{c}{\bf Density} \\
			\multicolumn{4}{c}{} & \multicolumn{1}{c}{$\cos\delta_T$} & \multicolumn{3}{c}{} \\
			\multicolumn{2}{c}{} & \multicolumn{1}{l}{$~h~~m~~s$}& \multicolumn{1}{l}{\,$~~^\circ~~~\prime~~~\prime\prime$} & \multicolumn{1}{c}{(\degrees)}  & \multicolumn{1}{c}{(\degrees)}& \multicolumn{1}{c}{(\degrees)}  & \multicolumn{1}{c}{(mJy)} & \multicolumn{1}{c}{(mJy)} \\
			\hline
			J0634--2335 && 06 34 59.00100 & --23 35 11.9573 & & &                    & 470 & $1100\pm200$\\
			& J0636--2113 &  06 36 00.60168 & --21 13 12.1997 &   0.2 &   2.3 & 2.38 & 200 & $210\pm35$\\
			& J0643--2451 &  06 43 07.46892 & --24 51 21.3120 &   0.9 & --1.3 & 2.25 & 130 & $170\pm30$\\
			& J0620--2515 &  06 20 32.11700 & --25 15 17.4851 & --3.3 & --1.7 & 3.69 & 320 & $400\pm70$\\
			& J0639--2141 &  06 39 28.72567 & --21 41 57.8045 &   1.0 &   1.9 & 2.15 & 130 & $40\pm8$\\
			& J0632--2614 &  06 32 06.50180 & --26 14 14.0353 & --0.7 & --2.7 & 2.73 & 230 & $630\pm100$\\
			& J0629--1959 &  06 29 23.76186 & --19 59 19.7236 & --1.3 &   3.6 & 3.82 & 750 & $970\pm150$\\\hline
			J1901--2112 && 19 01 04.45397 & --21 12 01.1656 & & &                    & 100 & $150\pm30$\\
			& J1916--1519 &  19 16 52.51100 & --15 19 00.0716 &   3.7 &   5.8 & 6.98 & 150 & $190\pm45$\\
			& J1848--2718 &  18 48 47.50417 & --27 18 18.0722 & --2.9 & --6.1 & 6.72 & 270 & $350\pm85$\\
			& J1928--2035 &  19 28 09.18336 & --20 35 43.7843 &   6.3 &   0.6 & 6.35 & 260 & $75\pm25$\\
			& J1832--2039 &  18 32 11.04649 & --20 39 48.2033 & --6.7 &   0.5 & 6.77 & 200 & $270\pm50$\\
			& J1916--2708 &  19 16 19.86268 & --27 08 32.2589 &   3.5 & --5.9 & 6.88 & 150 & $105\pm25$\\\hline
			J1336--0829 && 13 36 08.25983 & --08 29 51.7986 & & &                    & 220 & $560\pm125$\\
			& J1354--0206 &  13 54 06.89532 & --02 06 03.1906 &   4.4 &   6.4 & 7.81 & 450 & $615\pm150$\\
			& J1351--1449 &  13 51 52.64960 & --14 49 14.5569 &   3.9 & --6.3 & 7.40 & 150 & $505\pm115$\\
			& J1312--0424 &  13 12 50.90123 & --04 24 49.8923 & --5.7 &   4.1 & 7.08 & 140 & $280\pm50$\\
			& J1406--0848 &  14 06 00.70186 & --08 48 06.8806 &   7.4 & --0.3 & 7.38 & 360 & $225\pm45$\\
			& J1305--1033 &  13 05 33.01504 & --10 33 19.4281 & --7.6 & --2.1 & 7.82 & 350 & $240\pm50$\\
			& J1406--0707 &  14 06 10.81372 & --07 07 02.3097 &   7.4 &   1.4 & 7.56 & 200 & $260\pm70$\\\bottomrule
		\end{tabular}
	\end{table*}

    \subsection{Sources \& Observations}
\begin{figure*}[ht]
        \centering
        \includegraphics[width=0.32\textwidth]{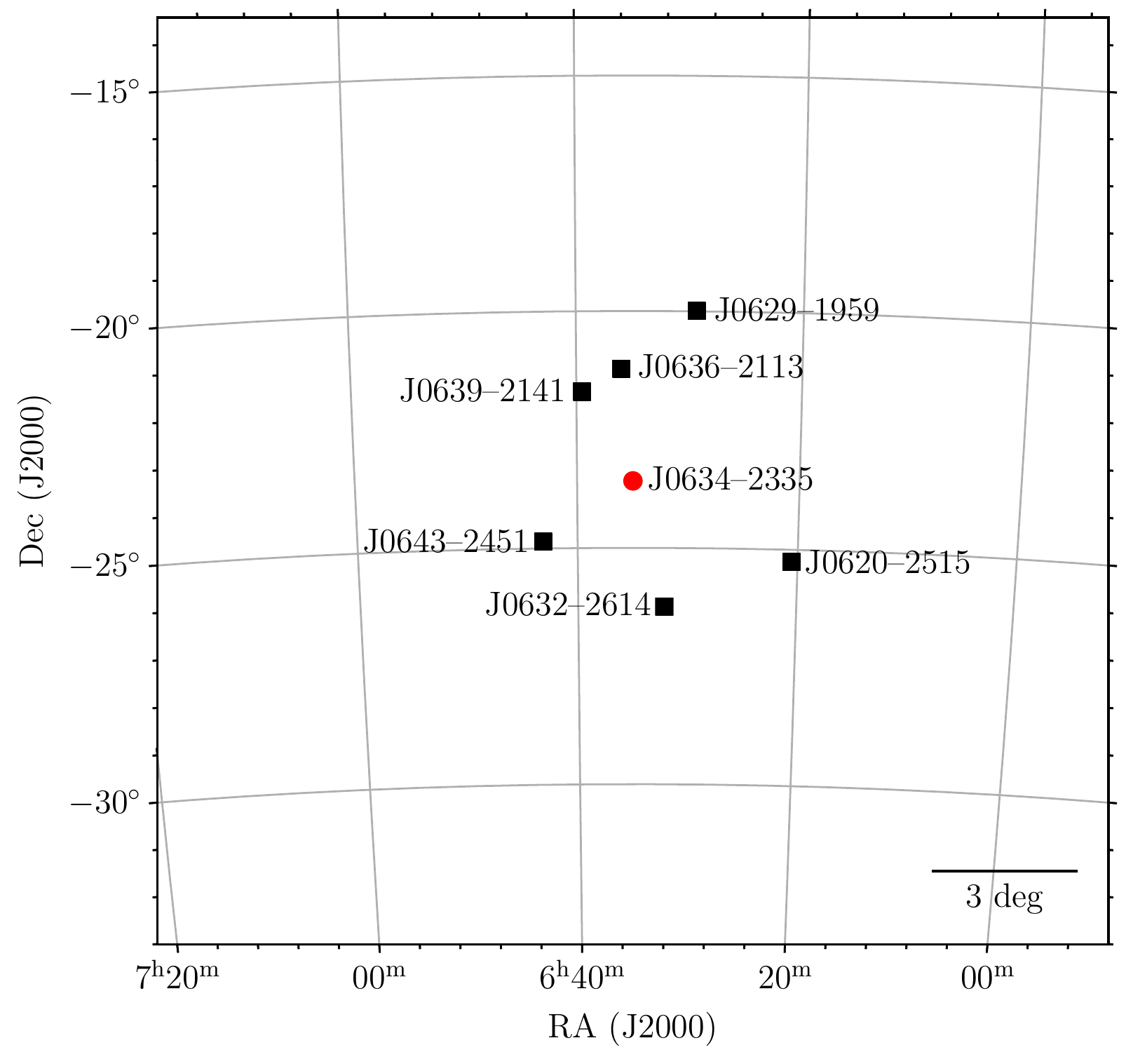}
        \includegraphics[width=0.32\textwidth]{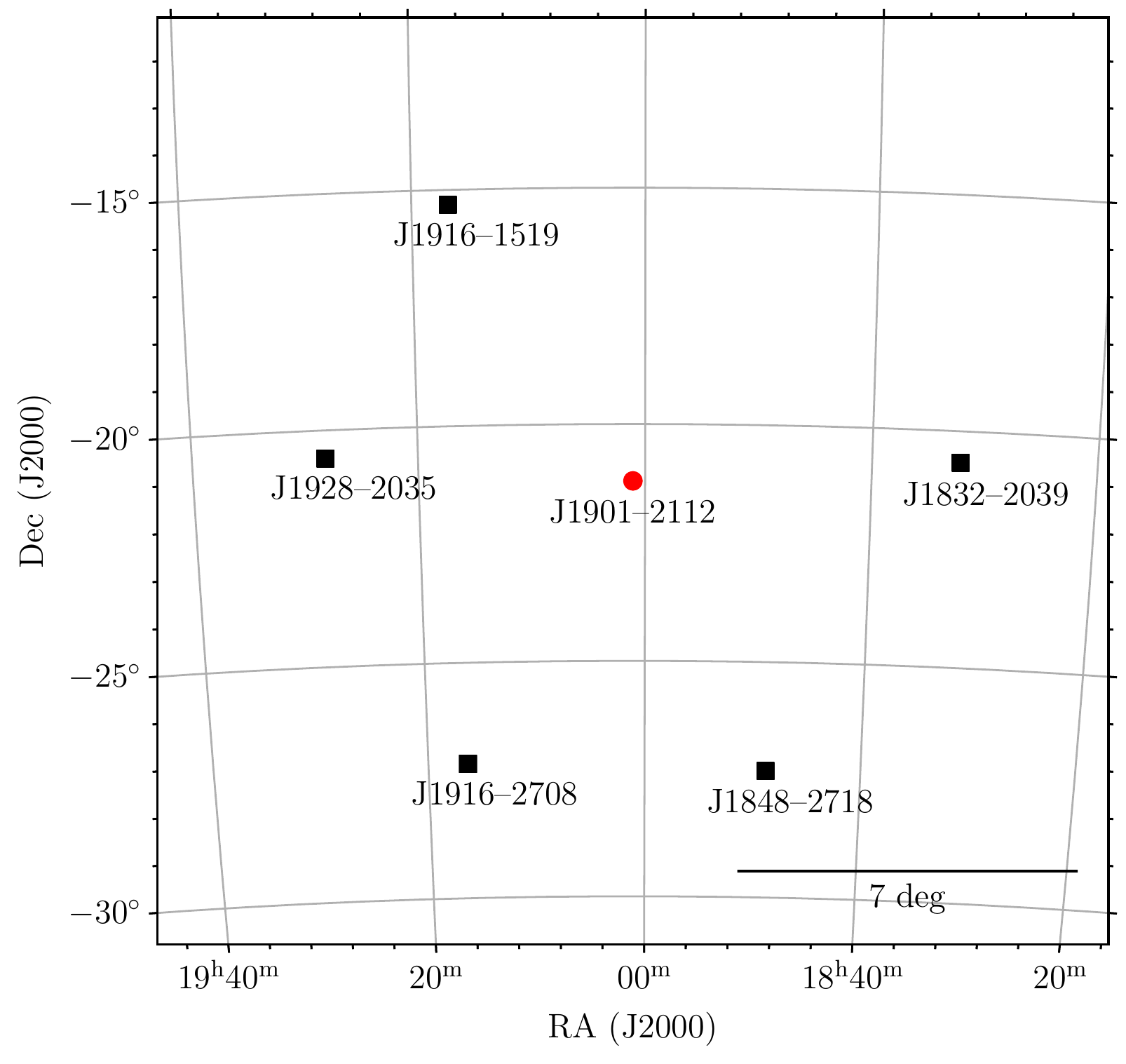}
        \includegraphics[width=0.32\textwidth]{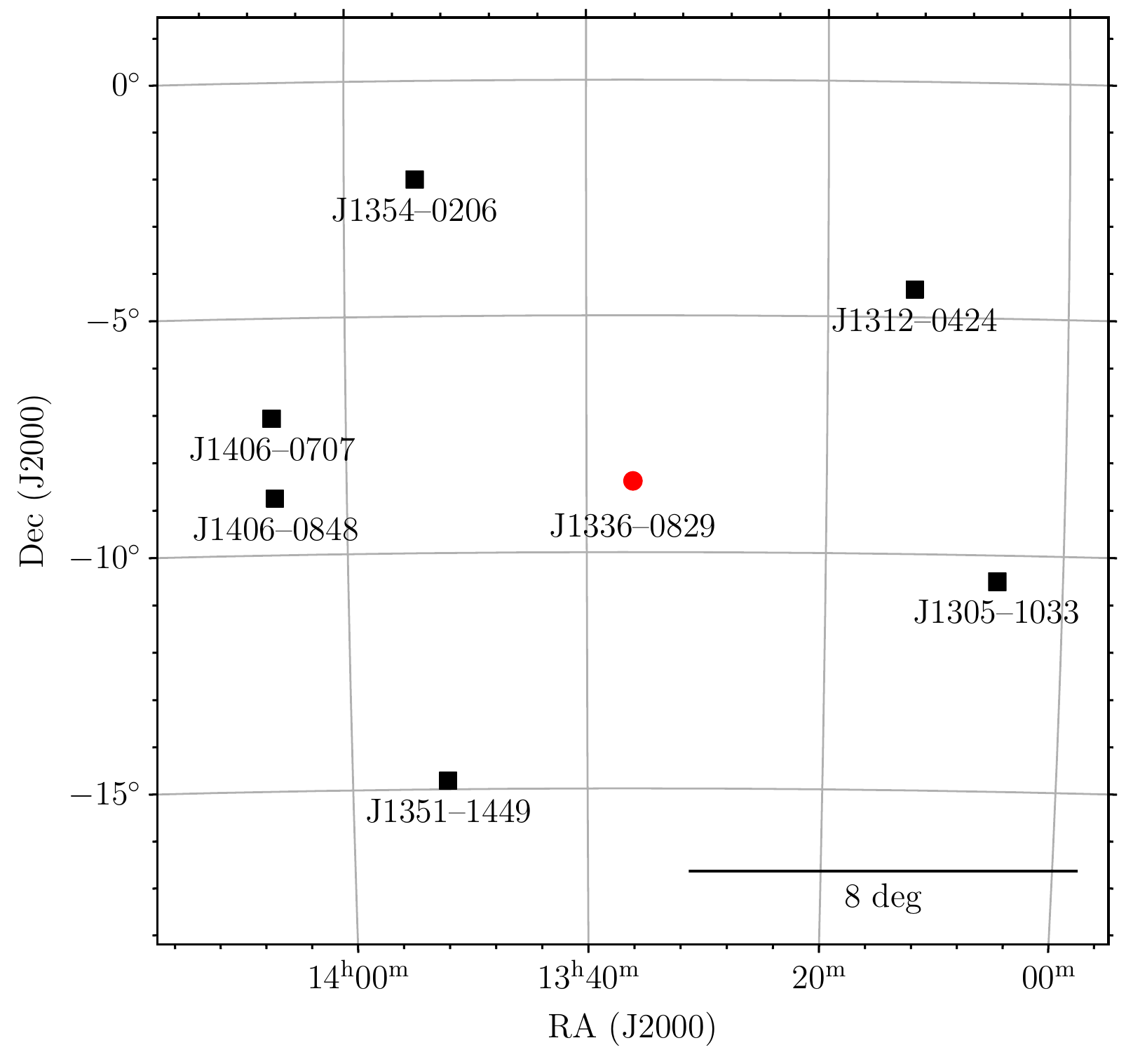}
        \caption{Sky distributions of the three clusters of quasars used for these tests. Targets (red circles) are surrounded by 5 or 6 calibrators (black squares) with mean separations (left to right) 2.8, 6.7 and 7.5\degrees.}
        \label{fig:quasarskydist}
    \end{figure*}

    We selected three groupings of quasars at right ascension 6, 13 and 19~hours with declination $<0$~\degrees~from the catalogue (rfc\_2019b) of \citet{Petrov2019}.  All quasars had a catalogued 8.3~GHz unresolved flux density $\ge100$~mJy (\Tab{tab:sources}) \red{and where possible, were chosen to have little to no structure}.  Five or six calibrator quasars were selected to be distributed in a thin `ring' around the target quasar (\Fig{fig:quasarskydist}) with mean radii of $\theta_\mathrm{sep} = 2.8, 6.7$ and $7.5$\degrees~respectively.
    
    Observations were constructed of 3 types of blocks: fringe-finder blocks (FFBs) for fringe-alignment and clock determination in the correlator, and preliminary electronic delay and phase calibration (i.e., manual phase calibration); geodetic-like blocks (Geoblock) for advanced delay calibrations; and iMV blocks.  These blocks were scheduled in the following repeated sequence: Geoblock, FFB, iMV block. 

    FFBs are comprised of 3-4 scans on strong ($\ge1$~Jy) quasars over a combined duration (on-source and slew) of $<15$~mins.  These quasars are optimally located close to the target to minimise slewing time and at $>45^\circ$ elevation at all telescopes.
    
    Geoblocks are short geodesy-style \citep{Heinkelmann13} observation periods that consist of $10-15$ ICRF2 quasars \citep{Fey1991} with sub-milliarcsecond accurate positions spread in elevation at each telescope site.  These 30~min blocks are scheduled every 3~hr and allow for correction of post-correlation residual tropospheric and clock delays at each telescope \citep{Honma2008,Reid2009,ReidHonma2014}. 
    
    The iMV blocks are a modified version of conventional inverse phase-referencing nodding sections \red{(\Eq{eq:iprnodding})} and are placed between the calibration blocks.  The iMV blocks form the primary data for the experiment. For a reference target ($T$) with $N$ calibrators ($C_{1\dots N}$), iMV blocks have the following sequence:
    \begin{equation}
        T, C_1, T, C_2, T, \dots, C_N, T, C_1, T, C_2, T, \dots, C_N, T
    \end{equation} 
    where target scans bracket sequentially different calibrator scans.  This allows all individual calibrators scans to be phase-referenced to the target in under the atmospheric coherence time ($\sim4$~min at 8.3~GHz). 
    
    The iMV blocks were $\sim150$~min duration, with each full cycle ($C_1\rightarrow C_N$) taking on average about 15~mins with individual on-source times of 50~sec, and the remaining time reserved for slewing. Therefore each $21$~hr observation contained seven calibration blocks and six iMV blocks, allowing two iMV blocks for each ring cluster spanning seven hours, which kept source elevations above $30^\circ$.
    
    In this way, we conducted four 21~hour observations of the three ring clusters on 2019 February 16, March 17, April 13 and May 4. \red{At these epochs the angular separation from the Sun was 120, 100, 85 and 70\degrees~for the smallest 2.8\degrees~ring; 45, 72, 100 and 120\degrees~for the 6.7\degrees~ring and; 120, 150, 175 and 160\degrees~for the largest 7.5\degrees~ring. The angular separation from the Sun is expected to influence the size of the ionospheric gradients encountered.}
    
    \subsection{Array \& Frequency}
\begin{figure}[ht]
    	\centering
    	\includegraphics[width=0.45\textwidth]{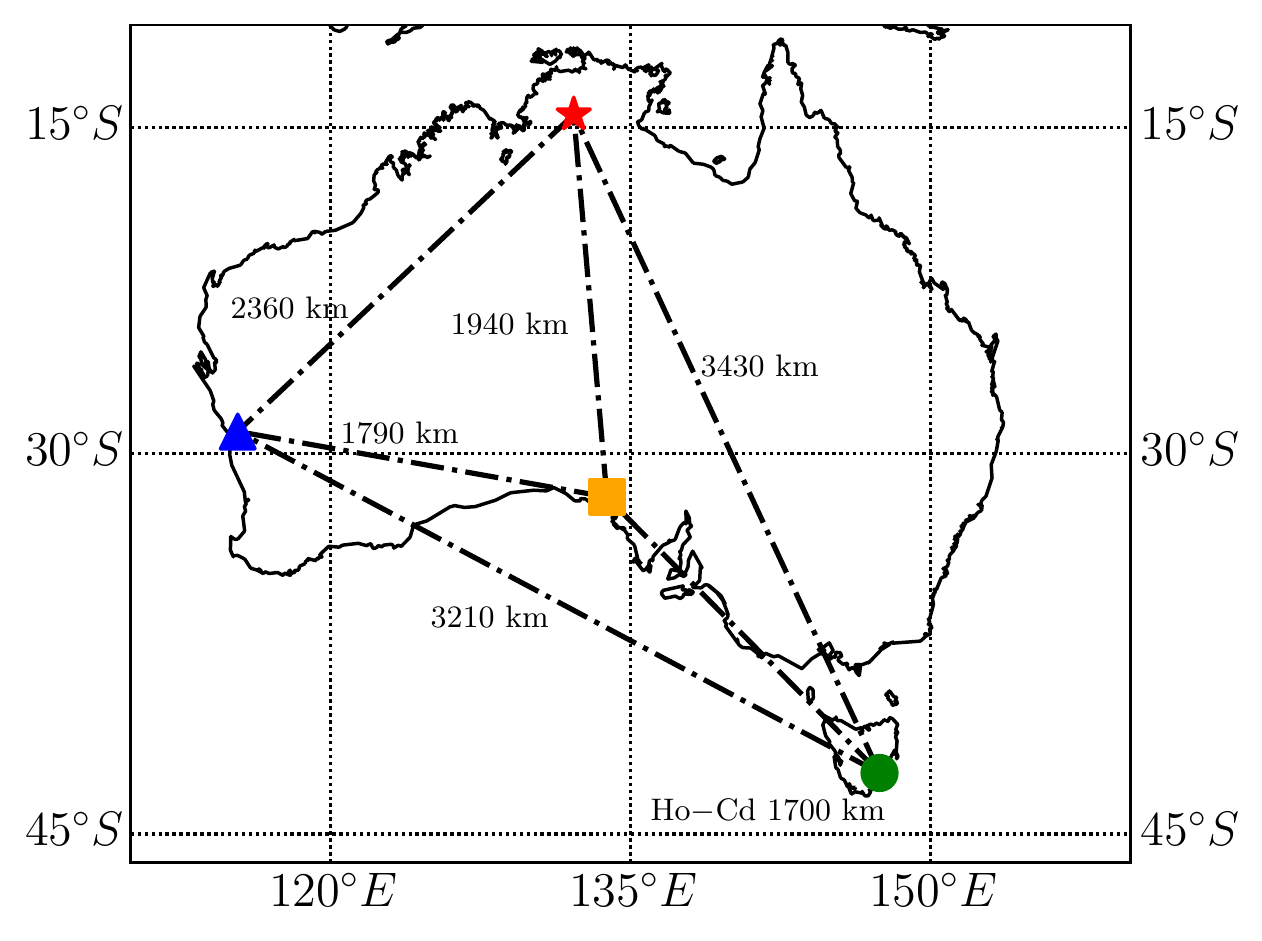}
    	\caption{The University of Tasmania VLBI array used for these observations includes the Ceduna~30m (Cd, orange square), Hobart~26m (Ho, green circle), Katherine~12m (Ke, red star) and Yarragadee~12m (Yg, blue triangle). \red{Geographic longitude and latitude are plotted on the $x$-axis and $y$-axis respectively.}}
    	\label{fig:array}
    \end{figure}

    Observations were conducted using the University of Tasmania AuScope-Ceduna Interfeometric (ASCI) Array \citep{Hyland2018}, comprising of Ceduna~30m \citep{McCulloch2005}, Hobart~26m, Katherine~12m and Yarragadee~12m \citep{Lovell2013}, with a maximum baseline of $|\textbf{B}|\approx3500$~km (\hyperref[fig:array]{Figure~\ref*{fig:array}}).
    
    Data were recorded at 1024~Mbps in right circular polarization covering frequencies between 8200 and 8456~MHz, with Nyquist sampling and 2-bits per sample.  Baseband data were correlated using the DiFX-2 software correlator \citep{Deller2011} at 0.5~MHz spectral resolution. Raw correlated FITS files can be provided upon request.
    
    \subsection{Preliminary Calibrations} \label{sec:calib}
    Data were analyzed in \aips \citep{Greisen1990,Greisen2003} using standard VLBI tools and with the assistance of the python wrapper software ParselTongue/Obit \citep{Kettenis2006}.
    Correlated FITS files were loaded into \aips using the task FITLD and data observed during off--source periods or windstows were flagged with task UVFLG. This generally included only a few seconds at the start of some scans, with the exception of a $\sim1$~hr period at the end of the second epoch where Ceduna~30m was windstowed. 
    Task ACCOR was used to correct amplitude in the cross-correlation spectra arising from \red{digitizer} sampler threshold errors and antenna system temperatures were applied with task ANTAB. 
    Updated Earth Orientation Parameters \citep{Seidelmann1982} were downloaded\footnote{gdc.cddis.eosdis.nasa.gov/vlbi/gsfc/ancillary/solve\_apriori/} and applied with the task CLCOR/EOP, and Total Electron Content (TEC) maps based on global positioning system data were downloaded\footnote{gdc.cddis.eosdis.nasa.gov/gnss/products/ionex/} and applied with task VLBATECR.
     
    Data observed in the geoblocks were separated and processed as follows:
        \begin{enumerate}
            \item The task FRING was run on a single FFB scan to fit single-band delay and phase (whilst zeroing the delay-rate); then the task CLCAL was used to apply this solution to all geoblock quasars. This removes the time-invariable delays between intermediate frequency bands.
            \item FRING was then used to fit multi-band delays and rates on all geoblock quasars. These data, along with the antenna elevations, were used to determine tropospheric zenith delays, clock offsets and drift rates at each telescope \red{using an external, DELZN--equivalent program \citep[described in][]{Reid2009a}}. These values were applied to the iMV data with the task CLCOR/ATMO.
        \end{enumerate}
        
	Data for each ring group in the iMV blocks were processed as follows: 
	    \begin{enumerate}
	    \setcounter{enumi}{2}
        \item A single FFB scan was used to remove the single-band delays and phases (as with step 1 on the geoblocks).
        \item The target quasar was fit for phase and rate using the task FRING, and the solutions were applied to both the target quasar and its ring calibrators using the task CLCAL.
        \item The calibrators were imaged using the task IMAGR, and their peak emissions were fitted with the task JMFIT. 
        \item The above process was repeated for all four epochs and the positions of the ring calibrators were shifted to the mean measured offset with the task CLCOR/ANTP. This minimizes possible phase wrapping (see \Sec{sec:phasewrap}).  After the position corrections were applied to the data, a second iteration of the same calibration process was undertaken (from step 3, skipping this step). 
        \item The data for the ring calibrators were averaged in frequency using the task SPLAT and a fringe fit for phase was performed using the task CALIB.  The task TBOUT was used to print out the CALIB phase solutions so that they could be used for iMV fitting.
        \item The target quasars were averaged in frequency with task SPLIT. 
        \end{enumerate}
    
    \red{At VLBI resolutions, many sources are somewhat resolved and this can lead to additional phase variations (structure phase). For sources that are partially resolved, preliminary self calibration may be used (between steps 3 and 4) to correct for the structure phase of each calibrator and target separately. We found that the targets and calibrators selected for these observations were sufficiently compact as to not require this correction.}

\section{Analysis} \label{sec:analysis}
\subsection{Phase Wraps} \label{sec:phasewrap}
    The initial correlation used the catalogue quasar positions and antenna location which have a reported accuracies of 0.3~mas and 1~cm respectively, sufficient to yield residual interferometric phase variations of $\lesssim120$\degrees~over a track (at 8.3~GHz with a 3500~km baseline). With this effect alone, phases should therefore remain in the same $-180$\degrees,$180$\degrees~wrap over the track. However, the residual tropospheric and ionospheric delays generally introduce phases which can vary many times this amount over a track. 
    
\begin{figure}[ht]
		\centering
		\includegraphics[width=0.5\textwidth]{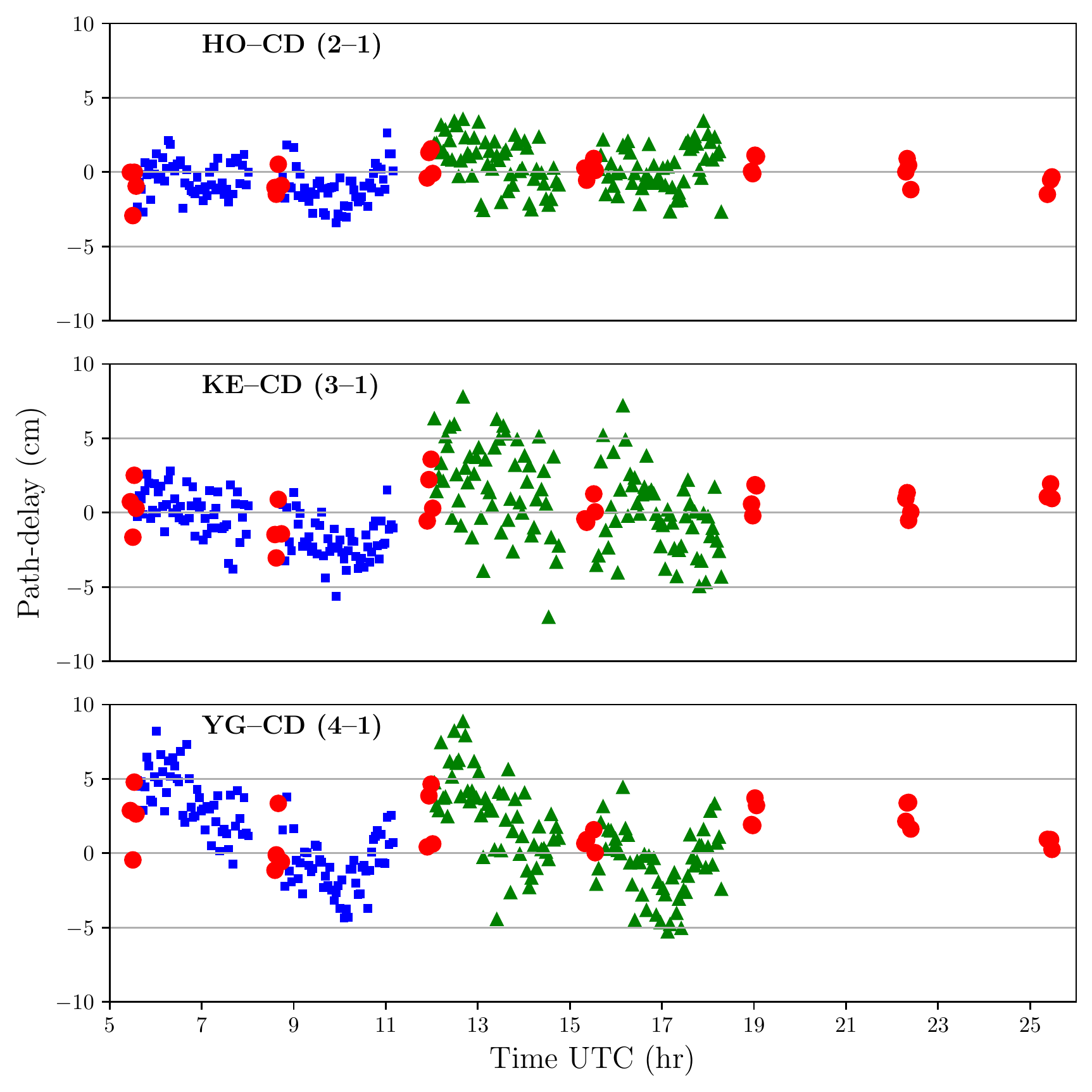}
		\caption{Residual \red{multi-band} path delay after preliminary calibration for select sources observed at epoch 3. The three panels show the three baselines to the reference antenna Ceduna~30m (Cd). \textbf{Colored markers:} Measured path delay for FFBs fringe-finders (red dots) and targets J0634-2335 (blue squares), J1336-0829 (green triangles). The final target, J1901-2112, is not shown as it was more than 6 times weaker than J0634-2335 or the fringe-finders observed in the FFBs after 19:00~UT. A fringe-finder in the FFB at 15:30~UT was used for the manual phase calibration at this epoch.}
		\label{fig:residualdel}
	\end{figure}

    Due to the abundance of relatively strong ($>300$~mJy) quasars in our data, we could measure the residual delay present after preliminary calibration (but before phase-referencing). We performed a fringe fit for multi-band delay (with task FRING) on target quasars and FFB quasars, and this revealed that there were residual delays in our data that had variations between 3 to 10~cm (\Fig{fig:residualdel}), equivalent to 300 to 1000\degrees~of phase.

    After phase referencing to the central target, the ring sources would then be expected to have {\it differential} phase errors of up to $\phi\approx1000\Delta\theta$\degrees, where $\Delta\theta$ is the angular separation in radians. For $\Delta\theta\approx0.1$~radians, we expect these atmospheric phase errors to approach 100\degrees, and the total amount of phase variation to possibly exceed 180\degrees. This can lead to some phase-wrapping \citep{RiojaDodson2020}, which need\red{s} to be corrected for before the iMV fitting process.
    
    In order to correct phase-wrapping, we minimised the absolute phase difference between consecutive scans on the same calibrator quasar, separated by approximately $\Delta t=15$~min (the average ring cluster duty-cycle), by adding or subtracting 360\degrees~as needed to keep the phase-difference,
    \begin{equation}
	    |\phi(t_{i+1}) - \phi(t_i)| < 180^\circ
    \end{equation}
    We used the phase measured at the middle of each track as a reference, since this is when the average antenna elevation is largest and we expect the phase differences due to the unmodeled residual atmospheric delays to be at a minimum.  For the data presented here, only a few baselines required correction for phase wrapping, rarely for the 2.8\degrees~ring, and more commonly for the larger separation clusters (1-2 times per track on average over all baselines).
    
\subsection{Inverse Multiview Fitting}
    We modelled the phase-screen over a quasar grouping as a 2D plane or `wedge' taking the form:
    \begin{equation}
        \phi_{i,jk} = \phi_{T,jk} + \mathcal{A}_{jk}~\Delta\alpha_i\cos\delta_T + \mathcal{B}_{jk}~\Delta\delta_i
        \label{eq:phaseplane}
    \end{equation} where $\Delta\alpha_i\cos\delta_T$ and $\Delta\delta_i$ are the angular offset from the target position for the $i^{th}$  calibrator (in degees on the sky).  The subscript $jk$ indicates an interferometer baseline.  The measured phase on the $i^{th}$ calibrator for the $jk$ baseline is given by $\phi_{i,jk}$ in degrees of phase, and $\phi_{T,jk}$ is the phase at the target position.  The parameters $\mathcal{A}_{jk}$, $\mathcal{B}_{jk}$ are the phase slopes in units of degrees of phase per degree of offset in the East-West and North-South directions, respectively.

\begin{figure}[ht]
		\centering
		\includegraphics[width=0.5\textwidth]{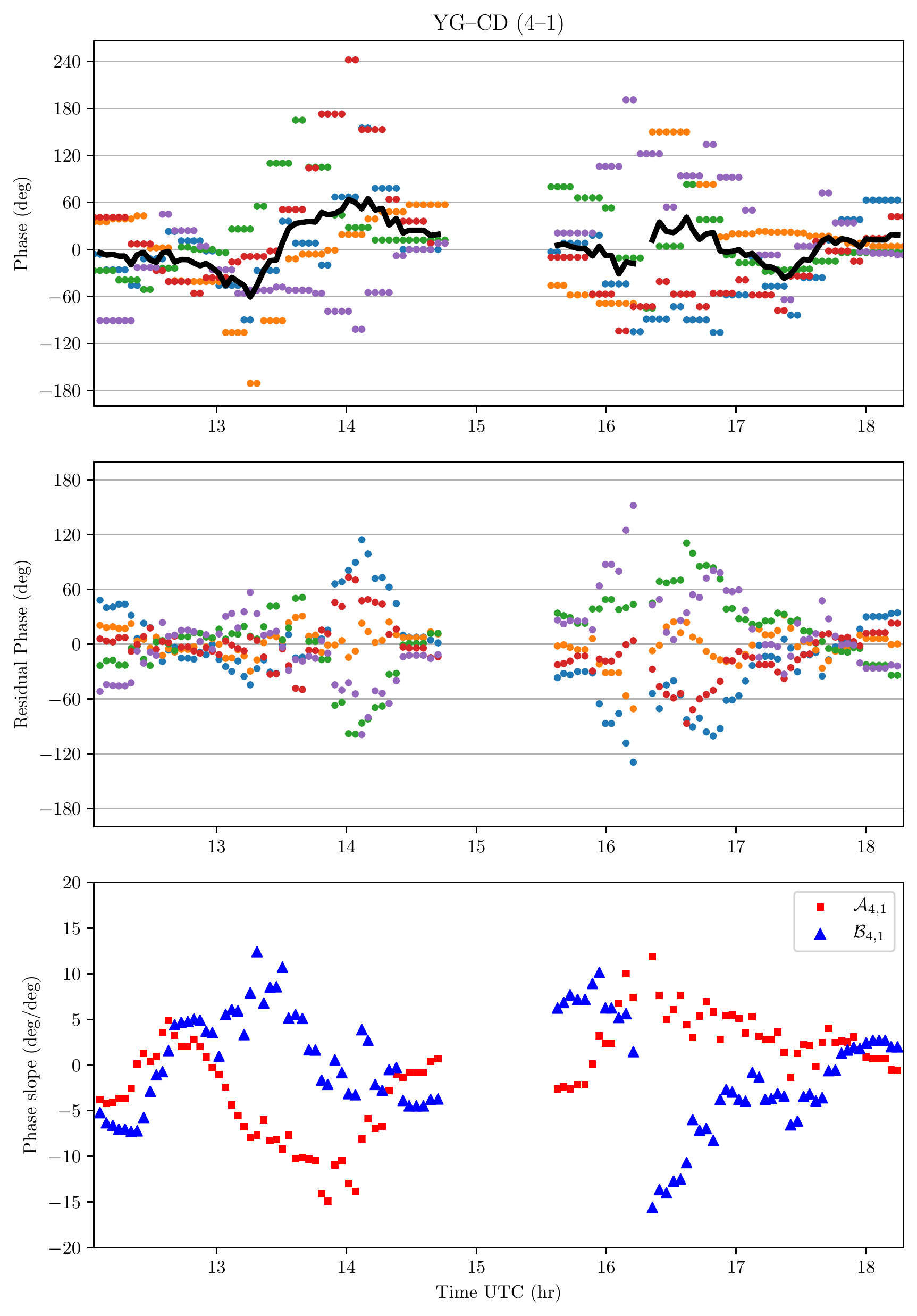}
		\caption{Inverse Multiview fitting for quasar cluster surrounding target source J1336-0829 with an average separation of 7.8\degrees~for the Yg--Cd baseline at epoch 3. \textbf{Top:} Unwrapped phases on each of 6 calibrator quasars (coloured dots) and fitted phase at target position (solid black line) over time. \textbf{Middle:} Residual phases after inverse Multiview fit. \textbf{Bottom:} Fitted phase slopes over time for the North-South (blue arrows) and East-West (red dots) directions respectively in degrees of phase per degree of separation.}
		\label{fig:imvfitting}
	\end{figure}

	Least-squares fitting was used to determine the phase slopes and target source phase in a sliding 15~min window interpolated to target quasar scan times (\Fig{fig:imvfitting}). As there will be a separate phase screen per antenna and all antennas have been referenced to a common antenna, we only solved \Eq{eq:phaseplane} for the $n-1$ baselines to that reference antenna (in this case $n=4$). The reference antenna was always chosen to be either Ceduna~30m or Hobart~26m.

    By comparing the residual phases and least-squares fit parameters over time, we were able to estimate if a quasar had an uncorrected phase wrap ambiguity; typically these occurred only once or twice per track. For example, in \Fig{fig:imvfitting} top panel, only the two red data points near 14:00~UT (which originally were at $-120^\circ$) and the two purple points at 16:30~UT (which originally were at $-170^\circ$) required correction for this effect.
    
    Finally, we flagged data where the mean absolute residuals exceeded 1~radian, as these indicated that the assumption of a planar phase gradient was not sufficient to explain the phase screen (for example, the break in solid black line at 16:\red{15}~UT in \Fig{fig:imvfitting} top panel). In general, only a few minutes of data was lost per track due to this, often near the beginning and/or end and accounting for less than 4\% of the total iMV block baseline hours observed.
	
    The target phase solutions, $\phi_{T,jk}$ (solid black line, top panel in \Fig{fig:imvfitting}), determined from the iMV process were loaded into \aips using the task TBIN to the SPLIT target data (from step 8  \red{in \Sec{sec:calib}}) as a solution (SN) table and then applied to the target data when imaging.
    Note that the phase-referencing during pre-iMV calibration had been applied to the target quasar (in addition to the ring quasars), leaving it with essentially zero phase.  This process effectively transfers all residual phase errors for the target quasar {\it equally} to all ring quasars, removing phase shifts due to both atmospheric and position errors.  However, the iMV phases, $\phi_{T,jk}$, which are defined at the position of the target quasar, will reflect phase shifts only associated with the target quasar's possible position shift, and the iMV ionospheric phase slopes are essentially side products.
    
    The target quasar was imaged using the task IMAGR and the peak emission in the images were fit with the task JMFIT.  Measured target quasar positions are given in \Tab{tab:positions}.
    Since quasars should have essentially zero (sub-microarcsecond) motions over our observations, we used the scatter in the sky positions $(x,y)$ (in $\mu$as) of the target quasars over the four epochs as the estimate for a single-epoch positional accuracy. In the East-West direction, the single-epoch positional accuracy ($\sigma_x$ in $\mu$as) was estimated with the standard deviation:
    \begin{equation}
        \sigma_x = \sqrt{\frac{1}{N-1}\sum_{i=1}^N (x_i - \overline{x})^2 }
        \label{eq:scatter}
    \end{equation} where $\overline{x}$ is the mean of $x$. The uncertainty in the single-epoch position accuracy ($SE_{\sigma}$) was estimated with:
    
    \begin{equation}
        SE_{\sigma_x} = \frac{\sigma_x}{\sqrt{2\left(N - 1\right)}}
    \end{equation} \citep[][]{Rao1973} where $N$ is the number of epochs. The single-epoch position accuracy and uncertainty in the North-South direction ($\sigma_y$,$SE_{\sigma_y}$) were estimated using analogous equations.

    \begin{figure}[ht]
        \centering
        \includegraphics[width=0.45\textwidth]{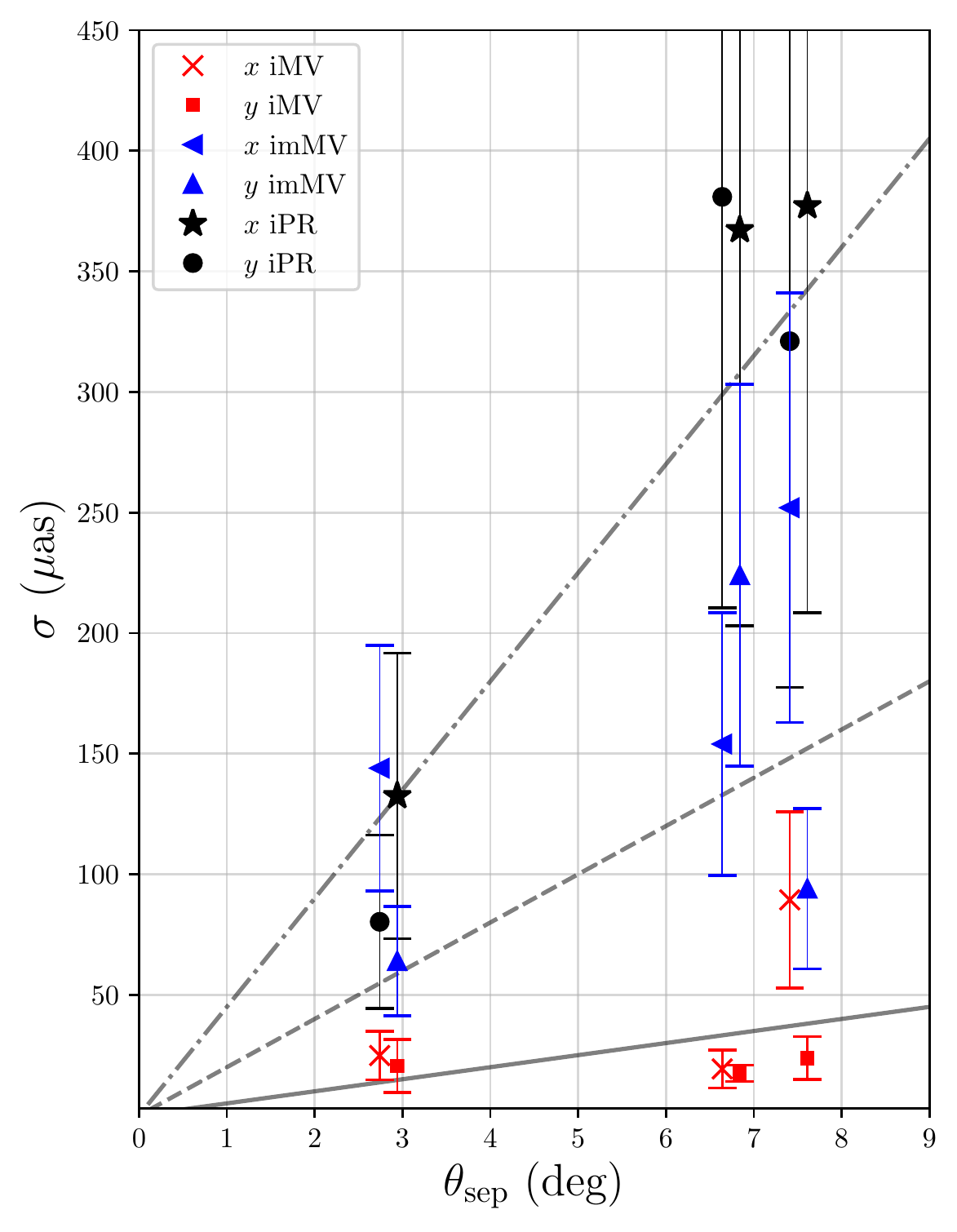}
        \caption{Astrometric accuracy vs. mean target-calibrator separation for the East--West ($x$) and North--South ($y$) directions after application of methods iMV,  ibMV, and iPR. Error bars show the 68\% CI. \textbf{Coloured markers:} Astrometric accuracy for iMV on target (red squares and crosses), for image-based Multiview (blue up- and left-facing triangles) and average of iPR (black stars and circles).     	
        \textbf{Black lines:} For comparison, the expected positional accuracy for iPR with $\sigma_\tau=0.5$~cm (solid), $\sigma_\tau=2$~cm (dashed) or $\sigma_\tau=4.5$~cm (dot--dashed) total residual path delay on a 3500~km baseline. Markers representing each coordinate have a small $x-$axis offset for clarity.}
        \label{fig:positionvsradius}
    \end{figure}

\subsection{Comparison with iPR and imaged-based Multiview}
        In order to compare iMV with iPR accuracy, we replicated the analysis process that would be undertaken if each calibrator was separately referenced to the target (e.g. \Eq{eq:iprnodding}). In such a case, each calibrator would be separately imaged, have its offset from the phase centre measured at each epoch, and then the (negative of the) offsets would be used over time to infer target apparent motion. The results from each calibrator could then be combined. Therefore, we took the scatter in offset for each calibrator over time (\Eq{eq:scatter}), then took the average of these scatters for each ring (black stars and circles in \Fig{fig:positionvsradius}).
    
        In the comparison with imaged-based Multiview, the calibrators are first separately referenced to the target, imaged and have offsets measured at each epoch (as with iPR).  Next, these offsets are used to construct an `artificial quasar' at the location of the target.  The motion of this artificial quasar should then mirror that of the target. Therefore we followed the ``Method-2" from \citet{Reid2017} -- fitting a tilted plane (with slopes $S_x,S_y$ and constants $C_x,C_y$) to each of the $x$ and $y$ offsets of the calibrator quasars at each epoch as a function of their angular separation from the target: 
        \begin{equation}
        \begin{split}
            x_i &= S_{x}\Delta\alpha_i\cos\delta + C_x \\ 
            y_i &= S_{y}\Delta\delta_i + C_y
        \end{split}
        \end{equation} where $\Delta\alpha_i\cos\delta_T$ and $\Delta\delta_i$ are the respective East-West and North-South angular separations for the $i^{th}$ calibrator quasar from the target at position $(\alpha_T,\delta_T)$ (\Tab{tab:sources}). The \red{slopes $S_x,S_y$ should reflect the average delay gradient at each epoch, and the} constants $C_x,C_y$ give the likely position shift of the targets at each epoch\red{. It is from these position shifts that} we determined the \red{imMV} scatter (blue triangles in \Fig{fig:positionvsradius}).       
        
\subsection{Thermal Uncertainty} \label{sec:therm}
    The thermal noise in an astrometric image limits how \red{precisely} the position of a feature in said image can be determined. We refer to this accuracy limit as the thermal uncertainty ($\sigma_\mathrm{pos,th}$), and estimate its magnitude as $\sigma_\mathrm{pos,th} \approx \red{0.5}~{\theta_B}/{\mathrm{SNR}}$ \red{\citep{Reid1988}}. Here $\theta_B$ is the synthesized beam (set by the ratio of the observing wavelength $\lambda$ to the baseline length $|\textbf{B}|$) and SNR is the signal-to-noise ratio for the feature in the images. 
    
    For the final astrometric images of the three targets, the average synthesized beam was $\theta_B = 2.3\times1.3$~mas at position angle 60\degrees~(\red{hereafter using the geometric mean} $\overline{\theta}_B = 1.73$~mas), and the average SNR was 175, 110, and 70 for the 2.8, 6.7, and 7.5\degrees~separation targets respectively. This gave a respective estimate for the thermal uncertainty as $\sigma_\mathrm{pos,th}=\red{5}$, \red{8} and \red{12}~$\mu$as for each of the three separations.

\section{Results \& Discussion} \label{sec:discussion}
	For iMV the single epoch accuracy is $\pm20~\mu$as in both coordinates for calibrator separations from the target of at least 6.7\degrees. As expected, for iPR the accuracies are separation dependent, growing from about $\pm100~\mu$as at $2.8$\degrees~to about $\pm300~\mu$as at 6.7\degrees.  The accuracies of imMV are between those of iPR and iMV, since imMV is, essentially, an observing-track average of a changing phase screen.
	
	There is one possible outlying measurement in our data in \Tab{tab:positions} at epoch 3 in the 7.5\degrees~ring (J1336--0829) for the East-West ($x$) direction.  Removing this measurement would yield a standard deviation of the remaining three points of $45\pm22~\mu$as, more in keeping with our other results.  One possible reason for having an outlier is an unresolved phase wrap ambiguity, which is expected to be more likely for larger target--calibrator separations.  Another potential source of problems for iMV calibration is tropospheric water-vapour fluctuations, which would not be expected to give a planar ``phase wedge" over the larger angular separation of some of the rings we observed, and in such cases would not be properly corrected by iMV. Experience with standard phase referencing at 22~GHz suggests significant degradation in image quality if a target is separated from a calibrator by more than about $3^\circ$ on the sky.  For a given (non-dispersive) delay error, interferometer phases scale linearly with observing frequency, and the $3^\circ$ ``limit" at 22~GHz scales to about $8^\circ$ at our observing frequency of 8.3~GHz. Even when including this possible outlier point, iMV still gives a much better single-epoch uncertainty than either imMV at $250~\mu$as or iPR at $325~\mu$as for the same coordinate and separation.
		
	These tests of the iMV technique show a dramatic improvement in accuracy for VLBI astrometry at 8.3~GHz compared to standard phase-referencing (\Fig{fig:positionvsradius}).  For example, \citet{Reid2017} found evidence for single-epoch position errors of $\approx100~\mu$as per degree of separation between a target and a calibrator at an observing frequency of 6.7~GHz. Scaled to an observing frequency of 8.3~GHz, this error estimate is close to the dot-dashed line in \Fig{fig:positionvsradius} inferred from our iPR results.
	
	Comparing the iMV results with those that would have been obtained from imMV imply that fitting in the visibility domain is preferable to fitting in the image domain, and gives a factor of 2-10 improvement in accuracy (\Fig{fig:positionvsradius}). This can be likely explained by the ability to measure and correct the for `spatially dynamic' components of the phase-screen that changes on hour-scales (e.g. $\mathcal{A}_{jk}(t)$, \Fig{fig:imvfitting}) whereas imMV effectively averages this effect over the whole track and whole array.
	
	The iMV astrometric results are consistent with a single-epoch accuracy near $\pm20~\mu$as, \red{which is approaching} the thermal uncertainty estimates from \Sec{sec:therm}. Furthermore, the iMV results appear largely separation independent, as would be expected if the accuracy were dominated by thermal processes \citep{RiojaDodson2020}. This is a dramatic improvement over standard phase referencing at this frequency, where typical residual ionospheric path-delays of $5$~cm would translate to roughly $\pm100~\mu$as (for a baseline length 3500~km and target-calibrator separation 2\degrees). 
	
	With accuracies near $20~\mu$as for iMV at 8.3~GHz using calibration sources with separations up to $\approx7^\circ$ separations, we have matched or exceeded some of the best results obtained at 22~GHz \citep[e.g.][]{Reid2019,vera2020} where ionospheric effects are about seven times smaller. As the \red{direct} Multiview analysis would be expected to have much the same performance as iMV, we expect that future observations will reveal that it achieves \red{comparable} astrometric errors\red{.}
	
\section{Concluding Remarks} \label{sec:conclusion}
	To date, the highest astrometric accuracies have been obtained from inverse phase referencing at 22~GHz using the VLBA, a homogeneous array of 10 antennas with a maximum baseline length of around 8500~km.  These benchmark results are obtained only for small ($<2$ \degrees) angular separations between the target and reference source, coupled with accurate ``geodetic-block'' modeling, where non-dispersive delays, owing mostly to water vapor in the atmosphere, dominate.  Residual dispersive delays at 22~GHz, after applying global TEC models, are generally small ($\sim1$~cm path delay).
	Here we have shown that through application of the new technique of inverse Multiview it is possible to obtain comparable astrometric accuracy for small arrays of heterogeneous antennas with shorter baseline lengths, at lower frequencies and in the presence of much larger residual delays. \red{Multiview methods will be central to achieving ultra-precision astrometry on the next generation of instruments, such as ngVLA and SKA \citep[for further reading, see sect. 7.2 in][]{RiojaDodson2020}.}

	The Bar and Spiral Structure Legacy (BeSSeL) Survey and the VERA project have measured parallaxes to $\approx250$ massive young stars which display water or methanol maser emission. This has resulted in a partial map of the spiral structure of the Milky Way.  We have begun parallax observations of 6.7~GHz methanol masers associated with massive young stars that cannot be seen with telescopes in the northern hemisphere in order to complete this map.  Astrometric results at this frequency using iMV, which we will explore in paper II, have achieved single-epoch accuracies \red{approaching} $20~\mu$as, consistent with the results of the tests documented here.  This should allow parallax accuracies of $\approx10~\mu$as, which translate to 10\% distance uncertainty at 10~kpc from the Sun.  

\section*{Data and code availability}
    The data underlying this article will be shared on reasonable request to the corresponding author.  The programmes and scripts used for data reduction and analysis are available from \github.

\section*{ACKNOWLEDGEMENTS}
		
\indent This research was supported by the Australian Research Council (ARC) Discovery Grant DP180101061. We want to thank Mr Brett Reid and Dr Warren Hankey for helping maintain and organising all University of Tasmania radio telescopes and Mrs Beverly Benson for managing the Ceduna~30m radio telescope. We acknowledge the Amangu, Jawoyn, Paredarerme and Wiriangu peoples as the traditional owners of the land situating the Yarragadee, Katherine, Hobart and Ceduna telescopes respectively. This research has made use of NASA’s Astrophysics Data System Abstract Service. This research made use of Astropy, a community-developed core Python package for Astronomy \citep{astropy:2013,astropy:2018}.
		
		
\bibliographystyle{yahapj}
\bibliography{multiview_paper}		


\begin{table}[ht]
    \caption{Measured positional offsets for the three target quasars over the four epochs and the determined positional accuracy. \textbf{Columns:} Average target-calibrator separation (1), target name (2), direction corresponding to data (3), measured position over four epochs (4-7), average positional offset from phase centre (8), single--epoch positional accuracy with 68\% CI (9).}
    \centering
    \begin{tabular}{cccrrrrrc}
    \toprule
\multicolumn{1}{c}{\bf Average} &
\multicolumn{1}{c}{\bf Target} &
\multicolumn{1}{c}{\bf Direction} &
\multicolumn{4}{c}{\bf Offset at Epoch} &
\multicolumn{1}{c}{\bf Mean} &
\multicolumn{1}{c}{\bf Scatter} \\
\multicolumn{1}{c}{\bf Separation} &
\multicolumn{2}{c}{} &
\multicolumn{1}{c}{\bf 1} &
\multicolumn{1}{c}{\bf 2} &
\multicolumn{1}{c}{\bf 3} &
\multicolumn{1}{c}{\bf 4} &
\multicolumn{1}{c}{\bf Offset} &
\multicolumn{1}{c}{$\boldsymbol{\sigma}$} \\
\multicolumn{1}{c}{(\degrees)} &
\multicolumn{2}{c}{} &
\multicolumn{1}{c}{ ($\mu$as)} &
\multicolumn{1}{c}{ ($\mu$as)} &
\multicolumn{1}{c}{ ($\mu$as)} &
\multicolumn{1}{c}{ ($\mu$as)} &
\multicolumn{1}{c}{ ($\mu$as)} &
\multicolumn{1}{c}{ ($\mu$as)} \\
        \midrule
        2.8&J0634$-$2335&EW&  69&  42&  16&  18 & $36$  & $25\pm10$\\
                         &&NS&  26& 29&  26&--14 & $17$  & $21\pm8$\\\hline
        6.7&J1901$-$2112&EW&--14&--57&--26&--19 & $-29$ & $19\pm8$\\
                         &&NS&  12&  13&  34&  48 & $27$  & $17\pm7$\\\hline
        7.5&J1336$-$0829&EW&--100&--15&--212&--32 & $-90$ & $89\pm37$\\
                         &&NS&--48&--52&--11&--67 & $-45$ & $24\pm10$\\\bottomrule
    \end{tabular}
    \label{tab:positions}
\end{table}

\end{document}